\documentclass[submission,copyright,creativecommons]{eptcs}
\providecommand{\event}{ACL2 2020} 
\usepackage{underscore}           

\usepackage{graphicx}
\usepackage{xcolor} 
\usepackage{amsmath}
\usepackage{amsfonts}
\usepackage{amssymb}
\usepackage{listings}
\definecolor{commentsColor}{rgb}{0.497495, 0.497587, 0.497464}
\definecolor{keywordsColor}{rgb}{0.000000, 0.000000, 0.635294} 
\definecolor{stringColor}{rgb}{0.558215, 0.000000, 0.135316}
\title{Quadratic Extensions in ACL2}
\author{Ruben Gamboa \qquad\qquad John Cowles
\institute{University of Wyoming \\ Laramie, Wyoming}
\email{\{ruben,cowles\}@uwyo.edu}
\and
Woodrow Gamboa
\institute{Stanford University \\ Stanford, California}
\email{woodrowg@stanford.edu}
}
\def\titlerunning{Quadratic Extensions}
\def\authorrunning{R. Gamboa, J. Cowles, \& W. Gamboa}
\begin{document}
\maketitle

\begin{abstract}
Given a field $K$, a quadratic extension field $L$ is an extension 
of $K$ that can be generated from $K$ by adding a root of a quadratic 
polynomial with coefficients in $K$. This paper shows how ACL2(r) can 
be used to reason about chains of quadratic extension fields 
$Q = K_0 \subseteq K_1 \subseteq K_2 \subseteq \cdots$, where each 
$K_{i+1}$ is a quadratic extension field of $K_i$. Moreover, we show that 
some specific numbers, such as $\sqrt[3]{2}$ and $\cos \frac{\pi}{9}$, cannot belong to any of the 
$K_i$, simply because of the structure of quadratic extension fields. 
In particular, this is used to show that $\sqrt[3]{2}$ and $\cos \frac{\pi}{9}$ are not rational.
\end{abstract}

\section{Introduction}

A field is a mathematical structure that supports addition, subtraction,
multiplication, and division in a way that satisfies the usual properties
of these operations in ordinary arithmetic \cite{gilbert2004modern}.
Fields can be made up of complicated objects (e.g., rational functions) with peculiar 
operations corresponding to addition and multiplication, but in this paper
we are concerned only with numeric fields, in which the objects in the field
are numbers and the operations are the very same ones from ordinary arithmetic.
Some common examples of numeric fields include the rationals $\mathbb{Q}$, the
reals $\mathbb{R}$, and the complex numbers $\mathbb{C}$. Notice that
$\mathbb{Q} \subseteq \mathbb{R} \subseteq \mathbb{C}$, and we say that
$\mathbb{R}$ is a field extension of $\mathbb{Q}$, and similarly
$\mathbb{C}$ is a field extension of $\mathbb{R}$ (and $\mathbb{Q}$).

It turns out that there are many field extensions that are intermediate between
$\mathbb{Q}$ and $\mathbb{R}$. One way to extend a given field $K$ is to start with
a number $x_1$ that is not already in $K$, then consider the closure of
$K\cup\{x_1\}$ under the typical arithmetic operators; the resulting field is 
called $K(x_1)$, and it is the smallest
numeric field that contains $x_1$ and all the elements of $K$. For example, we can
extend $\mathbb{Q}$ by adding the irrational number $\sqrt{2}$. The resulting field
$\mathbb{Q}(\sqrt{2})$ contains numbers such as 3, 2/7, and -12 (which were already
in $\mathbb{Q}$), $\sqrt{2}$ (which is explicitly added), and more involved numbers,
such as $\frac{(3-\sqrt{2})\sqrt{2}}{\sqrt{2}+5}$. It is clear that 
$\mathbb{Q} \subsetneq \mathbb{Q}(\sqrt{2})$. Although it may not be immediately
clear, it is also true that $\mathbb{Q}(\sqrt{2}) \subsetneq \mathbb{R}$. For instance,
$\sqrt[3]{2} \not\in \mathbb{Q}(\sqrt{2})$.

The process of extending $\mathbb{Q}$ by an irrational number can be repeated.
Let $K_0 = \mathbb{Q}$. Then a field $K_i$ can be extended by finding a $x_{i+1}$ that
is the root of a quadratic polynomial with coefficients from $K_i$ and letting
$K_{i+1}= K_i(x_{i+1})$. For example, starting with $K_0=\mathbb{Q}$, we can define
$K_1 = K_0(\sqrt{2})$ since $\sqrt{2}$ is a root of the polynomial $x^2-2$ which
has rational coefficients. Then we can define $K_2 = K_1(\frac{\sqrt{2}+\sqrt{6}}{2})$,
since $\frac{\sqrt{2}+\sqrt{6}}{2}$ is a root of the polynomial $x^2 - \sqrt{2}x -1$ with
coefficients in $K_1=\mathbb{Q}(\sqrt{2})$. Repeating this process indefinitely
results in a tower of quadratic field extensions $\mathbb{Q}=K_0 \subsetneq K_1
\subsetneq K_2 \subsetneq \cdots$. The main result in this paper is a formal proof
in ACL2(r) that for any such tower of quadratic field extensions where all the $x_i$ are real, 
$\cup_{i=0}^\infty K_i \subsetneq \mathbb{R}$. In particular, 
$\sqrt[3]{2} \not\in \cup_{i=0}^\infty K_i$, which immediately shows that
$\sqrt[3]{2}$ is irrational.

Before proceeding to the details of the ACL2 formalization, it may be helpful to
pause and explain our interest in these towers of field extensions. We are interested
in formalizing the impossibility of certain geometric constructions with straight-edge
and compass, such as trisecting an angle. Such constructions consist of arbitrarily
choosing two points, then drawing lines between points and circles centered
about a point and with a  radius defined by the distance
between two points, and finding more points by the intersection of such
lines and circles. The original arbitrary point and distance can be called 0 (the origin)
and 1 by fiat. The key to the proof of impossible constructions is that each step in
the construction process discovers a new point by solving a linear or quadratic equation,
i.e., the intersection of two lines, two circles or a line and a circle. So starting with 
$K_0$, the field generated by 0 and 1 (which happens to be $\mathbb{Q})$, we can construct
a tower of quadratic field extensions $\mathbb{Q}=K_0 \subsetneq K_1
\subsetneq K_2 \subsetneq \cdots$. Each step in
a straight-edge and compass construction results in points whose coordinates must be
the roots of a polynomial with coefficients in the previous extension field. For example,
starting with $\mathbb{Q}=K_0$ which contains the points 0 and 1, a geometric construction 
can find a line passing through 1, perpendicular to the line through 0 and 1, and then a point on 
that perpendicular that is precisely a unit from 1.
In Cartesian coordinates, the original points are at $(0,0)$ and $(0,1)$, and the newly
discovered point is at $(1,1)$. This defines the length between $(0,0)$ and the new
point $(1,1)$, which is easily seen to be $\sqrt{2}$. Thus, $\sqrt{2}$ can be constructed
using straight-edge and compass, and indeed $\sqrt{2} \in Q(\sqrt{2})$ which is in a 
tower of quadratic extensions of $\mathbb{Q}$. However, since any number that can be 
constructed using straight-edge and compass must be in some tower of quadratic extensions,
numbers like $\sqrt[3]{2}$ which \textbf{cannot} belong to any such tower also cannot be
the result of any straight-edge and compass construction, no matter how clever. In particular,
this shows that it is impossible to ``double a cube,'' i.e., to construct using only straight-edge
and compass a cube with twice the volume of another cube, since doubling a cube with volume 1
requires constructing a cube with side $\sqrt[3]{2}$, which is impossible from the informal
discussion in this paragraph. Likewise, trisecting an angle is impossible, since trisecting a
$\frac{\pi}{3}$ angle results in a length of $\cos \frac{\pi}{9}$, which we also show in ACL2 is not 
constructible; i.e., $\cos \frac{\pi}{9}$ cannot belong to any tower of quadratic field 
extensions\footnote{Another famous impossible construction is that of squaring a circle, i.e.,
finding a square with the same area as a given circle. This can be shown using the fact that
$\pi$ is transcendental and hence not in any tower of field extensions, since the side of a square 
with the same area as the unit circle must have length $\sqrt{\pi}$. We are currently working
on formalizing this fact in ACL2.}.

The remainder of this paper is structured as follows. Section~\ref{sec:field-props} introduces
the basic notion of fields and towers of fields in ACL2(r). Section~\ref{sec:field-ext-is-field}
shows that the elements in the field $K(\sqrt{k})$ when $k\in K$ but $\sqrt{k}\not\in K$, can all 
be written as $a + b\sqrt{k}$ for unique $a,b\in K$. This key property is used to show that
extending a field $K$ simply by using linear combinations involving $\sqrt{k}$ and elements of $K$
actually results in the field $K(\sqrt{k})$. Section~\ref{sec:field-ext-polys} introduces
polynomials. In particular, it shows that if $a+b\sqrt{k}$ is a root of a polynomial with 
coefficients in $K$, then so is $a-b\sqrt{k}$. That is, roots come in conjugate pairs. This is
then used to show that cubic polynomials with coefficients in a given $K_i$ that have at least
one root in $K_{i+1}$ must also have a root in $K_i$. Thus, cubic polynomials with rational
coefficients with a root in any $K_i$ must also have at least one rational root. This section
also shows a proof of the Rational Root Theorem, which can be used to list all possible rational
roots of a polynomial with rational coefficients. This theorem is then used to show that some
rational cubic polynomials cannot have any rational roots (since none of the finite possible candidate
roots are in fact roots of the polynomial), and therefore that these cubic polynomials cannot have any
roots in any tower of quadratic field extensions. Since $\sqrt[3]{2}$ and $\cos \frac{\pi}{9}$ are
roots of such polynomials, they cannot belong to any such tower and must be irrational.
Finally, Section~\ref{sec:conclusions} concludes the paper by discussing ongoing and future work.

\section{Basic Field Properties}
\label{sec:field-props}

We formalize the notion of \emph{numeric field} in ACL2 with a constrained function
\texttt{number-field-p} that recognizes elements of a (generic) field. The constraints 
on this function enforce the following:
\begin{itemize}
\item Any element of the field is a number, possibly complex.
\item Both 0 and 1 must belong to any field.
\item The field is closed under arithmetic operations.
\end{itemize}
There is no need to include the typical ``field axioms'' for the operations, since numeric 
fields use the ordinary arithmetic operators, and ACL2 already knows that the ordinary 
arithmetic operators always satisfy the field axioms.
We note in passing that it follows directly from these constraints that $Q \subseteq K$ is true
for any numeric field $K$, and this was easily verified in ACL2.

Consider a field $K$ and its extension by $x_1$. It is in fact the case that all elements in
$K(x_1)$ can be written as $a + b x_1$, for some $a, b \in K$. Now consider extending $K(x_1)$ 
by introducing $x_2$, resulting in the field $K(x_1, x_2) \supsetneq K(x_1)$. As before, an 
arbitrary element of $K(x_1, x_2)$ can be written as $a' + b' x_2$, for $a', b' \in K(x_1)$.
But since both $a'$ and $b'$ can be written as $a + b x_1$ for some choice of $a, b \in K$,
it follows that each element of $K(x_1, x_2)$ can be written as $a + b x_1 + c x_1 + d x_1 x_2$ for
some $a, b, c, d \in K$. This pattern continues for extensions by a finite number of points
$x_1$, $x_2$, \dots, $x_n$, and we use this pattern to define towers of extensions, since this
definition is much more concrete and amenable to ACL2 than a direct translation of ``all numbers
that result from finite applications of the arithmetic operators to the elements
in $K\cup\{x_1\}$.'' For the rest of this paper, the original field $K$ is fixed as $\mathbb{Q}$.

We formalize this in ACL2 with a handful of functions. First is \texttt{eval-linear-combination} 
which takes in a set of ``coordinates'' (e.g., $a$ and $b$) and a ``spanning set'' 
(e.g., $1$ and $\sqrt{2}$), and returns their dot product (i.e., $a + b\sqrt{2}$). Another
useful function is \texttt{all-products} which takes in a list and returns a list of the 
products of subsets of the original list. For example, if the input list is $\langle \sqrt{2}, 
\sqrt{3}, \sqrt{5} \rangle$, \texttt{all-products} will return $\langle 1, \sqrt{5}, \sqrt{3}, 
\sqrt{3}\sqrt{5}, \sqrt{2}, \sqrt{2}\sqrt{5}, \sqrt{2}\sqrt{3}, 
\sqrt{2}\sqrt{3}\sqrt{5} \rangle$. Finally, there is the important function \texttt{is-linear-combination-p}
that recognizes members of $\mathbb{Q}(x_1, x_2, \dots, x_n)$. This is defined in ACL2 as follows:
\begin{lstlisting}
(defun-sk is-linear-combination-p (x exts)
  (exists coords
          (and (rational-listp coords)
               (equal (len coords) (expt 2 (len exts)))
               (equal (eval-linear-combination coords 
                                               (all-products exts))
                      x))))
\end{lstlisting}
We note that the argument \texttt{exts} contains the list of $x_i$ extending $\mathbb{Q}$ but in
reverse order. I.e., to check whether $x \in \mathbb{Q}(x_1, x_2, \dots, x_n)$, we would use the
ACL2 expression:
\begin{itemize}
\item \texttt{(is-linear-combination-p x '(x_n \dots\ x_2 x_1))}
\end{itemize}
This reversal of the natural order is common in ACL2 code, because of the asymmetry of list processing
with \texttt{car}, \texttt{cdr}, and \texttt{cons}.

At this point, we have \emph{syntax} for recognizing members of a quadratic field extension
$\mathbb{Q}(x_1, x_2, \dots, x_n)$, but we have not yet shown that our recognizer actually works correctly,
since we are using the indirect notion of linear combination instead of the direct notion of field
extension in the recognizer. It should be obvious that any element admitted by our recognizer really
does belong to $\mathbb{Q}(x_1, x_2, \dots, x_n)$, but it is possible that not all elements of
$\mathbb{Q}(x_1, x_2, \dots, x_n)$ are properly recognized. This issue can be resolved if we show that
the members recognized by \texttt{(is-linear-combination-p x exts)} form a mathematical field (with
suitable conditions on \texttt{exts}). We do so in the following section.

\section{Quadratic Field Extensions Are Field Extensions}
\label{sec:field-ext-is-field}

Suppose we know that the elements recognized by \texttt{(is-linear-combination-p x '(x_n \dots\ x_2 x_1))}
correspond to $\mathbb{Q}(x_1, x_2, \dots, x_n)$. We want to consider what happens when we add a new
number $x_{n+1}$. But first, let's quickly dispense with the case of removing the point $x_n$. It should
be obvious that if $x$ is a linear combination of the products of $\langle x_1, x_2, \dots, x_{n-1} \rangle$,
then it is also a linear combination of the products of $\langle x_1, x_2, \dots, x_n \rangle$.
In ACL2, we have
\begin{lstlisting}
(defthm is-linear-combination-is-tower
  (implies (consp exts)
           (implies (is-linear-combination-p x (cdr exts))
                    (is-linear-combination-p x exts)))
  :hints ...)
\end{lstlisting}
This theorem is important, because it establishes the fact that the structure recognized by
the predicate \texttt{is-linear-combination-p} is a tower of enclosing sets. It remains to be 
shown that it is a tower of field extensions.

Let $S_n$ be the set recognized by \texttt{(is-linear-combination-p x '(x_n \dots\ 
\mbox{x_2} \mbox{x_1}))}. Note that if $S_n$ is in fact a field, then it must be the smallest field
that contains $\mathbb{Q}$ and the elements $x_1$, $x_2$, \dots, $x_n$,
i.e., it must be $\mathbb{Q}(x_1, x_2, \dots, x_n)$. 
This is true, since each element in $S_n$ is a linear combination with rational coefficients
of products of the $x_i$, and since addition and multiplication are field operations, each
element of $S_n$ must be in the field $\mathbb{Q}(x_1, x_2, \dots, x_n)$. 
In other words, $S_n \subset \mathbb{Q}(x_1, x_2, \dots, x_n)$. But then, 
$\mathbb{Q}(x_1, x_2, \dots, x_n)$
is the smallest field containing all these elements, so if $S_n$ happens to be a field, 
it must also be that $\mathbb{Q}(x_1, x_2, \dots, x_n) \subset S_n$.
Note that this argument justifies the
definition of $\mathbb{Q}(x_1, x_2, \dots, x_n)$ in ACL2 using \texttt{is-linear-combination-p}.
However, this argument is necessarily a paper-and-pencil proof and not formalized in ACL2, since
the set $\mathbb{Q}(x_1, x_2, \dots, x_n)$ is not explicitly defined in ACL2 other than 
using \texttt{is-linear-combination-p}.

What is done in ACL2 is to show that the sets $S_n$ do form a numeric field. It is immediately clear
that if the $x_i$ are numbers, so is any element of $S_n$. Moreover, both $0$ and $1$ (and in fact
all rationals) are in $S_n$. The only non-trivial property is that $S_n$ is closed under arithmetic
operations.

To show that $S_n$ is closed under addition, we need to show that $x+y$ is a linear combination of
a spanning set, given that both $x$ and $y$ are. This is easily done by considering the component-wise
sum of the coordinates of $x$ and $y$. In ACL2, we have
\begin{lstlisting}
(defthm sum-of-linear-combinations
  (implies (equal (len coords1) (len coords2))
           (equal (eval-linear-combination (add-coords coords1 coords2)
                                           exts)
                  (+ (eval-linear-combination coords1 exts)
                     (eval-linear-combination coords2 exts)))))

(defthm is-linear-combination-p-is-closed-addition
  (implies (and (is-linear-combination-p x exts)
                (is-linear-combination-p y exts))
           (is-linear-combination-p (+ x y) exts))
  :hints ...)
\end{lstlisting}
Here \texttt{add-coords} simply adds corresponding coordinates.

Similarly, we can show that $S_n$ is closed under additive inverses by simply negating the
coordinates of $x$. In ACL2, this results in
\begin{lstlisting}
(defthm is-linear-combination-p-has-additive-inverse
  (implies (is-linear-combination-p x exts)
           (is-linear-combination-p (- x) exts))
  :hints ...)
\end{lstlisting}
These last two theorems are simple, and typical of proofs in ACL2.

Multiplication, however, is more complicated. Conceptually, it is similar to addition and negation,
but the algebra is considerably more complicated. For one thing, in the case of addition the structure
of the spanning set was completely irrelevant. If $x = a + b\alpha$ and $y = c + d\alpha$, then
$x+y = (a+c) + (b+d)\alpha$, regardless of the value of $\alpha$. But consider $x\cdot y
= ac + ad\alpha + bc\alpha + bd\alpha^2$. This cannot be written in the form $A + B\alpha$ unless there is 
something special about $\alpha^2$. This is why we need to have some special requirements on the
elements $x_1$, $x_2$, \dots, $x_n$ that make up the spanning set \texttt{exts} in
\texttt{is-linear-combination-p}. At a minimum, we should have that $x_n^2$ is a linear combination
of the product of the $x_1$, $x_2$, \dots, $x_{n-1}$; this will take care of the the $\alpha^2$ term
above. It is also helpful to insist that $x_n$ is \emph{not} a linear combination
of the product of the $x_1$, $x_2$, \dots, $x_{n-1}$. Although this is not strictly necessary, it
means that each extension actually does extend the field in some way. It will also become important
later, when we use it to show that the coordinates of a linear combination are actually unique. In
ACL2, this is captured with the following definition:
\begin{lstlisting}
(defun quadratic-extensions-p (exts)
  (if (consp exts)
      (and (acl2-numberp (first exts))
           (not (is-linear-combination-p (first exts) (rest exts)))
           (is-linear-combination-p (expt (first exts) 2) (rest exts))
           (quadratic-extensions-p (rest exts)))
    (equal exts nil)))
\end{lstlisting}
It remains only to show how the product of two elements in
$\mathbb{Q}(x_1, x_2, \dots, x_n)$ must be in $\mathbb{Q}(x_1, x_2, \dots, x_n)$.
The key fact is that since $\mathbb{Q}(x_1, x_2, \dots, x_n)$ is an extension of
$\mathbb{Q}(x_1, x_2, \dots, x_{n-1})$, any element in the former
can be written as $a + b x_n$ where $a,b \in \mathbb{Q}(x_1, x_2, \dots, x_{n-1})$. So if we have two elements of $\mathbb{Q}(x_1, x_2, \dots, x_n)$
their product can be written as 
$$(a_1 + b_1 x_n)(a_2 + b_2 x_n) = 
 a_1 a_2 + a_1 b_2 x_n + a_2 b_1 x_n + b_1 b_2 x_n^2
 = (a_1 a_2 + b_1 b_2 x_n^2) + (a_1 b_2 + a_2 b_1) x_n$$
This is in the form $a + b x_n$ since $x_n^2\in\mathbb{Q}(x_1, x_2, \dots, x_{n-1})$,
hence so $a_1 a_2 + b_1 b_2 x_n^2$. In ACL2, this is formalized as follows, 
where the \texttt{(take ...)} and \texttt{(nthcdr ...)} expressions above
serve to find the coefficients $a_i$ and $b_i$ in $a_i + b_i x_n$:
\begin{lstlisting}
(defthm eval-linear-combination-product-split-1
  (implies (and (equal (len coords1) (expt 2 (len exts)))
                (rational-listp coords1)
                (equal (len coords2) (expt 2 (len exts)))
                (rational-listp coords2)
                (acl2-number-listp exts))
           (equal (* (eval-linear-combination coords1 
                                              (all-products exts))
                     (eval-linear-combination coords2 
                                              (all-products exts)))
                  (if (consp exts)
                      (+ (* (eval-linear-combination 
                              (take (expt 2 (len (rest exts))) coords1)
                              (all-products (rest exts)))
                            (eval-linear-combination 
                              (take (expt 2 (len (rest exts))) coords2)
                              (all-products (rest exts))))
                         (* (first exts)
                            (eval-linear-combination 
                              (take (expt 2 (len (rest exts))) coords1)
                              (all-products (rest exts)))
                            (eval-linear-combination 
                              (nthcdr (expt 2 (len (rest exts))) 
                                      coords2)
                              (all-products (rest exts))))
                         (* (first exts)
                            (eval-linear-combination 
                              (take (expt 2 (len (rest exts))) coords2)
                              (all-products (rest exts)))
                            (eval-linear-combination 
                              (nthcdr (expt 2 (len (rest exts))) 
                                      coords1)
                              (all-products (rest exts))))
                         (* (expt (first exts) 2)
                            (eval-linear-combination 
                              (nthcdr (expt 2 (len (rest exts))) 
                                      coords1)
                              (all-products (rest exts)))
                            (eval-linear-combination 
                              (nthcdr (expt 2 (len (rest exts))) 
                                      coords2)
                              (all-products (rest exts)))))
                    (* (first coords1) (first coords2)))))
  :hints ...)
\end{lstlisting}
This serves to justify the following function which explicitly finds the coefficients
of the linear combination of the product:
\begin{lstlisting}
(defun product-coords (coords1 coords2 exts)
  (if (consp exts)
      (append (add-coords (product-coords 
                            (take (expt 2 (len (rest exts))) coords1)
                            (take (expt 2 (len (rest exts))) coords2)
                            (rest exts))
                          (product-coords 
                            (is-linear-combination-p-witness 
                              (expt (first exts) 2) 
                              (rest exts))
                            (product-coords 
                              (nthcdr (expt 2 (len (rest exts))) 
                                      coords1)
                              (nthcdr (expt 2 (len (rest exts))) 
                                      coords2)
                              (rest exts))
                            (rest exts)))
              (add-coords (product-coords 
                            (take (expt 2 (len (rest exts))) coords1)
                            (nthcdr (expt 2 (len (rest exts))) coords2)
                            (rest exts))
                          (product-coords 
                            (take (expt 2 (len (rest exts))) coords2)
                            (nthcdr (expt 2 (len (rest exts))) coords1)
                            (rest exts))))
    (list (* (first coords1) (first coords2)))))
\end{lstlisting}
Note how \texttt{is-linear-combination-p-witness} is used in this definition to find the
coefficients of $x_n^2$ in $\mathbb{Q}(x_1, x_2, \dots, x_{n-1})$.

What follows is a tedious (but not terribly illuminating) algebraic proof  that 
\texttt{product-coord} does in fact capture the product of its two arguments. This culminates 
in the following theorem:
\begin{lstlisting}
(defthm product-of-linear-combinations
  (implies (and (quadratic-extensions-p exts)
                (equal (len coords1) (expt 2 (len exts)))
                (rational-listp coords1)
                (equal (len coords2) (expt 2 (len exts)))
                (rational-listp coords2))
           (equal (eval-linear-combination (product-coords coords1 
                                                           coords2 
                                                           exts)
                                           (all-products exts))
                  (* (eval-linear-combination coords1
                                              (all-products exts))
                     (eval-linear-combination coords2
                                              (all-products exts)))))
  :instructions ...)
\end{lstlisting}
Once this theorem is proven, it is trivial to show that \texttt{is-linear-combination-p} is
closed under multiplication:
\begin{lstlisting}
(defthm is-linear-combination-p-is-closed-multiplication
  (implies (and (quadratic-extensions-p exts)
                (is-linear-combination-p x exts)
                (is-linear-combination-p y exts))
           (is-linear-combination-p (* x y) exts))
  :hints ...)
\end{lstlisting}

To show that \texttt{is-linear-combination-p} is also closed under division
we continue with the observation that any element $z$ in 
$\mathbb{Q}(x_1, x_2, \dots, x_n)$ can be written as $z = a + b x_n$ 
where $a,b \in \mathbb{Q}(x_1, x_2, \dots, x_{n-1})$. The ACL2 functions
\texttt{subfield-part} and \texttt{extension-part} extract these coefficients
$a$ and $b$:
\begin{lstlisting}
(defun subfield-part (x exts)
  (if (consp exts)
      (eval-linear-combination (take (expt 2 (len (rest exts)))
                                     (is-linear-combination-p-witness 
                                       x exts))
                               (all-products (rest exts)))
    (fix x)))
(defun extension-part (x exts)
  (if (consp exts)
      (eval-linear-combination (nthcdr (expt 2 (len (rest exts)))
                                       (is-linear-combination-p-witness 
                                         x exts))
                               (all-products (rest exts)))
    0))
\end{lstlisting}
To find $\frac{1}{z}=\frac{1}{a+b x_n}$, we employ a strategy common from
complex analysis. First we define the conjugate of $z$ as 
$\overline{z} = a - b x_n$. It follows then that 
$z \overline{z} = (a + b x_n) (a - b x_n) = a^2 - b^2 x_n^2$ must be in
$\mathbb{Q}(x_1, x_2, \dots, x_{n-1})$, since all of $a$, $b$, and $x_n^2$ are.
Thus,
$\frac{1}{z} = \frac{{\overline{z}}}{z{\overline{z}}} = 
\frac{a - b x_n}{z\overline{z}} = 
\frac{a}{z\overline{z}} - \frac{b}{z\overline{z}} x_n$, 
and this must be in $\mathbb{Q}(x_1, x_2, \dots, x_{n})$ since both
$\frac{a}{z\overline{z}}$ and $\frac{-b}{z\overline{z}}$ are in
$\mathbb{Q}(x_1, x_2, \dots, x_{n-1})$. There is a nagging detail, however.
What if $z \overline{z} = 0$, in which case multiplying by 
$\frac{{\overline{z}}}{{\overline{z}}}$ does not work as expected?
It turns out that this is an impossibility, unless $z = 0$. That is because
$\overline{z} = a - b x_n$, so this is $0$ only when
$a = b x_n$, in which case either $a=b=0$, so that $z=0$, or $b\ne0$ and
$x_n = \frac{a}{b}$. But this cannot be, since 
$x_n \not\in \mathbb{Q}(x_1, x_2, \dots, x_{n-1})$ whereas both
$a$ and $b$ are in $\mathbb{Q}(x_1, x_2, \dots, x_{n-1})$.

The proof of this in ACL2 follows
this outline, although the algebra is considerably tedious, and mostly involves
reasoning about which expressions are in $\mathbb{Q}(x_1, x_2, \dots, x_{n-1})$.
The end result is the following theorem, which completes the proof that
\texttt{is-linear-combination-p} recognizes a field, which must be exactly
$\mathbb{Q}(x_1, x_2, \dots, x_{n})$ :
\begin{lstlisting}
(defthm is-linear-combination-p-has-multiplicative-inverse
  (implies (and (quadratic-extensions-p exts)
                (is-linear-combination-p x exts)
                (not (equal x 0)))
           (is-linear-combination-p (/ x) exts))
  :hints ...)
\end{lstlisting}

\section{Quadratic Field Extensions and Polynomials}
\label{sec:field-ext-polys}

We will now explore how quadratic extension fields relate to roots of certain 
polynomials, and we begin this exploration with conjugates. But to fully explore
conjugates, it helps to show that the representation $x = a + b\omega$ for
$x\in K(\omega)$ where $a,b\in K$ but $\omega \not\in K$ is unique.

As is often the case, the key to the uniqueness theorem is to show that 0 is
unique. Indeed, if $0 = a + b\omega$, then $b$ must be 0. Otherwise,
$\omega=-a/b \in K$, which contradicts the assumption that $\omega \not\in K$.
But then, $0 = a + 0\omega = 0$. Thus $a + b\omega = 0$ implies that $a=b=0$.

Now suppose that $x = a_1 + b_1 \omega = a_2 + b_2 \omega$, where $a_i,b_i\in K$
but $\omega \not\in K$. Then 
$0 = a_1-a_2 + (b_1 - b_2)\omega$, so $a_1=a_2$ and $b_1 = b_2$. In ACL2, we have
\begin{lstlisting}
(defthmd subfield-extension-parts-unique
  (implies (and (consp exts)
                (quadratic-extensions-p exts)
                (is-linear-combination-p x exts)
                (is-linear-combination-p alpha (cdr exts))
                (is-linear-combination-p beta (cdr exts))
                (equal (+ alpha (* beta (car exts))) x))
           (and (equal (subfield-part x exts) alpha)
                (equal (extension-part x exts) beta)))
  :hints ...)	
\end{lstlisting}

Using this uniqueness theorem, it is straightforward to prove many properties
of conjugation. For example, let $x_1=a_1 + b_1 \omega$ and $x_2=a_2 + b_2 \omega$.
Then $x_1 + x_2 = a_1+a_2 + (b_1 + b_2)\omega$ and
$\overline{x_1 + x_2} = a_1+a_2 - (b_1 + b_2)\omega = 
a_1 - b_1 \omega + a_2 - b_2 \omega = \overline{x_1} + \overline{x_2}$.
Similarly, $\overline{x_1 - x_2} = \overline{x_1} - \overline{x_2}$. It is
also possible to show that $\overline{x_1 \cdot x_2} = \overline{x_1} \cdot \overline{x_2}$,
but that is slightly less direct. 
$x_1 \cdot x_2 = a_1 b_1 + a_1 b_2 \omega + a_2 b_1 \omega + a_2 b_2 \omega^2 
= (a_1 b_1 + a_2 b_2 \omega^2) + (a_1 b_2 + a_2 b_1)\omega$, remembering that $\omega^2 \in K$ 
if $K(\omega)$ is a quadratic extension (per our ACL2 definition). But then
$\overline{x_1} \cdot \overline{x_2} = 
a_1 b_1 - a_1 b_2 \omega - a_2 b_1 \omega + a_2 b_2 \omega^2 
= (a_1 b_1 + a_2 b_2 \omega^2) - (a_1 b_2 + a_2 b_1)\omega = \overline{x_1 \cdot x_2}$.
From this, it also follows that for non-zero $x$, $\overline{1/x}=1/\overline{x}$, since
$\overline{x} \cdot \overline{1/x} = \overline{x \cdot 1/x} = 1$.

Having proved the product rule for conjugates, a straightforward induction shows that
$\overline{x^n} = \overline{x}^n$. The product rule again, coupled with the fact that
$\overline{a}=a$ for any constant $a \in K$, can then be used to show that for any
monomial $\overline{a x^n} = a \overline{x}^n$. Another induction generalizes to any
polynomial $P$ with coefficients in $K$: $P(\overline{x}) = \overline{P(x)}.$ In
particular, if $x_0$ is a root of the polynomial $P$, then so is $\overline{x_0}$,
since $P(\overline{x_0}) = \overline{P(x_o)} = \overline{0} = 0$.
This important theorem tells us that roots come in conjugate pairs, and it will play
a major role in the sequel. In ACL2, it is written as
\begin{lstlisting}
(defthmd conjugate-of-root-is-root-of-polynomial
  (implies (and (quadratic-extensions-p exts)
                (is-linear-combination-p x exts)
                (is-linear-combination-listp poly (rest exts))
                (equal (eval-polynomial poly x) 0))
           (equal (eval-polynomial poly (qef-conjugate x exts)) 0))
  :hints ...)	
\end{lstlisting}

We now apply this theorem to the special case of cubic polynomials. Fix 
$P(x) = a_3 x^3 + a_2 x^2 + a_1 x + a_0$, where all $a_i \in K$ and
$a_3 \ne 0$. Suppose that $x_0$ is a root of $P$ such that $x_0 \in K(\omega)$ but $x_0 \not\in K$. From the
previous theorem, $\overline{x_0}$ is also a root of $P$. In fact,
with a bit of algebra, $P$ can be factored as
$$a_3 x^3 + a_2 x^2 + a_1 x + a_0 = a_3 (x - x_0) (x - \overline{x_0}) \left(x + \frac{a_2 + a_3 (x_0 + \overline{x_0})}{a_3}\right).$$
This shows that $C = - \frac{a_2 + a_3 (x_0 + \overline{x_0})}{a_3}$ is the remaining root of the cubic $P$.
But since $x_0 + \overline{x_0} \in K$ and all the coefficients $a_i \in K$, this shows that 
$C \in K$ also. In other words, if there is some root of $P$ that is in the extension
field $K(\alpha)$, there must be a (possibly different) root $x_1$ that is in $K$.
In particular, if $P$ has rational coefficients and there is a root $x_0$ of $P$ such that
$x_0 \in Q(x_1, x_2, \dots, x_n)$, then there is a (possibly different) root of $P$ in $x_0 \in Q(x_1, x_2, \dots, x_{n-1})$, and by induction there must also be a
root of $P$ that
is rational. In ACL2, we prove this as follows:
\begin{lstlisting}
(defthmd 
  poly-coeffs-in-subfield-and-root-in-field-implies-exists-rational-root
  (implies (and (quadratic-extensions-p exts)
                (polynomial-p poly)
                (rational-listp poly)
                (equal (len poly) 4)
                (not (equal (fourth poly) 0))
                (exists-root-in-field-extension poly exts))
           (exists-rational-root poly))
  :hints ...)	
\end{lstlisting}
Naturally, the \texttt{exists-*} functions are defined using \texttt{defun-sk}, e.g.,
\begin{lstlisting}
(defun-sk exists-rational-root (poly)
  (exists (x)
          (and (equal (eval-polynomial poly x) 0)
               (rationalp x))))
\end{lstlisting}

Of course, not all polynomials with rational (or even integer) coefficients have rational roots; 
$x^2 - 2$ is a famous counterexample. The Rational Roots Theorem from high school algebra can be
used to enumerate all the possible rational roots of a polynomial with integer
coefficients. Although this theorem applies to arbitrary integer polynomials,
we proved it in ACL2 only for cubic polynomials.

In particular, suppose $P(x) = a_3 x^3 + a_2 x^2 + a_1 x + a_0$ with 
$a_i\in\mathbb{Z}$ and
$P(p/q)=0$ for some rational $p/q$ in lowest terms. It follows that
$a_3 (p/q)^3 + a_2 (p/q)^2 + a_1 (p/q) + a_0 = 0$, and multiplying both sides
by $q^2$ yields
$\frac{a_3}{q} p^3 + a_2 p^2 + a_1 p q + a_0 q^2 = 0$. Since
$a_2 p^2 + a_1 p q + a_0 q^2$ is an integer, this implies that
$\frac{a_3 p^3}{q}$ is an integer, so $q$ divides $a_3$ (since $p$ and $q$ were chosen to be
relatively prime). Similarly, multiplying both
sides by $q^3/p$ gives
$a_3 p^2 + a_2 p q + a_1 q^2 + \frac{a_0}{p} q^3 = 0$. Again, this
can be used to conclude that $\frac{a_0}{p}$ is an integer, since all the other
terms are\footnote{Careful readers may notice that this makes use of the fact
that $\frac{0}{0}=0$ is well-defined in ACL2.}. This is proved in ACL2 with the following theorems:
\begin{lstlisting}
(defthmd rational-root-theorem-part-1
  (implies (and (polynomial-p poly)
                (integer-listp poly)
                (equal (len poly) 4)
                (rationalp x)
                (equal (eval-polynomial poly x) 0))
           (integerp (/ (fourth poly) (denominator x))))
  :hints )
(defthmd rational-root-theorem-part-2
  (implies (and (polynomial-p poly)
                (integer-listp poly)
                (equal (len poly) 4)
                (rationalp x)
                (equal (eval-polynomial poly x) 0))
           (integerp (/ (first poly) (numerator x))))
  :hints ...)
\end{lstlisting}

What this means is that the only possible rational roots of the polynomial
$P(x) = a_3 x^3 + a_2 x^2 + a_1 x + a_0$ must be of the form
$p/q$ where $q$ divides $a_3$ and $p$ divides $a_0$. In other words, factoring
$a_3$ and $a_0$ is sufficient to find all possible rationals that could be
roots of $P(x)$, and since this is a finite set, we can systematically consider 
all possible rational roots of $P(x)$. In some cases, of course, none of the
candidate rational roots will actually be roots of $P(x)$, so we can conclude
that $P(x)$ has no rational roots at all. And using the previous theorem, that
also means that $P(x)$ has no roots in any quadratic extension
$Q(x_1, x_2, \dots, x_n)$.

For example, consider the cubic polynomial $x^3-2$ with integer coefficients.
According to the Rational Root Theorem, the only rationals that could be
roots of this polynomial are $2$, $1$, $-1$, and $-2$:
\begin{lstlisting}
(defconst *poly-double-cube*   '(-2  0 0 1))
(defthmd possible-rational-roots-of-double-cube
  (implies (and (rationalp x)
                (equal (eval-polynomial *poly-double-cube* x) 0))
           (or (equal x 2)
               (equal x 1)
               (equal x -1)
               (equal x -2)))
  :hints ...)
\end{lstlisting}
A simple computation suffices to show that none of these candidates are actually
roots of the polynomials. Hence, we can conclude that the polynomial has no
rational roots:
\begin{lstlisting}
(defthm no-rational-roots-of-double-cube
  (implies (rationalp x)
           (not (equal (eval-polynomial *poly-double-cube* x) 0)))
  :hints ...)
\end{lstlisting}
Moreover, since this polynomial has no rational roots, it cannot have any roots
in any quadratic extension of $\mathbb{Q}$. In ACL2, this is proved as follows:
\begin{lstlisting}
(defthmd roots-not-in-quadratic-extension-double-cube
  (implies (and (quadratic-extensions-p exts)
                (equal (eval-polynomial *poly-double-cube* x) 0))
           (not (is-linear-combination-p x exts)))
  :hints ...)
\end{lstlisting}
Of course, the polynomial does have some roots, such as $\sqrt[3]{2}$. What
this means is that $\sqrt[3]{2}$ must not be rational and cannot belong to
any quadratic extension of $\mathbb{Q}$. We proved this in ACL2 as follows:
\begin{lstlisting}
(defthmd cube-root-of-two-is-root-of-poly-double-cube
  (equal (eval-polynomial *poly-double-cube* (raise-to 2 1/3)) 0)
  :hints ...)
(defthm cube-root-of-two-is-not-in-quadratic-extension
  (implies (quadratic-extensions-p exts)
           (not (is-linear-combination-p (raise-to 2 1/3) exts)))
  :hints ...)
(defthm cube-root-of-two-is-irrational
  (and (realp (raise-to 2 1/3))
       (not (rationalp (raise-to 2 1/3))))
  :hints ...)
\end{lstlisting}

A similar argument is sufficient to show that $\cos \frac{\pi}{9}$ is irrational.
Specifically, we applied the Rational Roots Theorem to the polynomial $8x^3-6x-1$,
and found the candidate roots $\pm 1/8$, $\pm 1/4$, $\pm 1/2$, and $\pm 1$.
None of these are actually roots of the polynomial, so we can conclude that 
it has no rational roots or roots in any quadratic extension field of
$\mathbb{Q}$. However, using the previously developed library of trigonometric
identities in ACL2, we showed that
$\cos(3x) = 4 \cos^3 x - 3\cos(x)$; hence,
$\frac{1}{2} = \cos(3\frac{\pi}{9}) = 4 \cos^3 \frac{\pi}{9} - 3\cos \frac{\pi}{9})$
and $\cos\frac{\pi}{9}$ is a root of $8x^3-6x-1$. Therefore, $\cos\frac{\pi}{9}$
must be irrational and not in any quadratic extension field of $\mathbb{Q}$:
\begin{lstlisting}
(defconst *poly-trisect-angle* '(-1 -6 0 8))
(defthmd roots-not-in-quadratic-extension-trisect-angle
  (implies (and (quadratic-extensions-p exts)
                (equal (eval-polynomial *poly-trisect-angle* x) 0))
           (not (is-linear-combination-p x exts)))
  :hints ...)
(defthmd cos-pi/9-is-root-of-trisect-angle
  (equal (eval-polynomial *poly-trisect-angle* 
                          (acl2-cosine (/ (acl2-pi) 9))) 
         0)
  :hints ...)
(defthmd cos-pi/9-not-in-quadratic-extension-trisect-angle
  (implies (quadratic-extensions-p exts)
           (not (is-linear-combination-p (acl2-cosine (/ (acl2-pi) 9)) 
                                         exts)))
  :hints ...)
(defthm cos-pi/9-is-irrational
  (and (realp (acl2-cosine (/ (acl2-pi) 9)))
       (not (rationalp (acl2-cosine (/ (acl2-pi) 9)))))
  :hints ...)
\end{lstlisting}

\section{Conclusions}
\label{sec:conclusions}

This paper formalized quadratic field extensions in ACL2, and it showed that 
certain numbers cannot belong to any quadratic field extension of $\mathbb{Q}$,
which also means those numbers must be irrational. This is all part of a larger
effort to formalize the notion of constructible numbers in ACL2, which leads
to the result that certain straight-edge and compass constructions are
impossible. 
For example, the facts that $\sqrt[3]{2}$ and $\cos\frac{\pi}{9}$
cannot belong to any quadratic field extension are the key to showing the 
impossibility of doubling a cube and trisecting an angle, respectively.
In the future, we plan to prove in ACL2 that $\sqrt\pi$ is also not constructible, 
since all constructible numbers are algebraic and $\sqrt\pi$ is not. This will
be used to show the impossibility of squaring the circle.

\nocite{*}
\bibliographystyle{eptcs}
\bibliography{quadratic-extensions}
\end{document}